\documentclass[prd,preprint,tightenlines,floatfix,showpacs,showkeys,date,preprintnumbers,nofootinbib,eqsecnum]{revtex4}
 \usepackage[dvips,final]{graphicx}
  \usepackage{bm}
   \usepackage{amsmath}
    \usepackage{amssymb}
     \usepackage{pifont}

\newcommand{\be}{\begin{equation}}\newcommand{\ee}{\end{equation}}%
\newcommand{\bd}{\begin{displaymath}}\newcommand{\ed}{\end{displaymath}}
\newcommand{\bit}{\begin{itemize}}                        
 \newcommand{\eit}{\end{itemize}}                         
\newcommand{\ben}{\begin{enumerate}}                      
 \newcommand{\een}{\end{enumerate}}                       
\newcommand{\baa}{\begin{array}{lll}}                     
 \newcommand{\eaa}{\end{array}}                           
\newcommand{\ba}{\begin{eqnarray}}                        
 \newcommand{\ea}{\end{eqnarray}}                         
  \newcommand{\footn}{\footnotesize}                      
\newcommand{\Ds}{\displaystyle}                           
\newcommand{\va}[1]{\langle{#1}\rangle}                   
\newcommand{\gev}[1]{\relax\ifmmode{\text{GeV}^{#1}}      
                     \else{GeV$^{#1}${ }}\fi}             
\newcommand{\Gev}{\relax\ifmmode{\text{GeV}}              
                     \else{GeV{ }}\fi}                    
\newcommand{\Mev}{\relax\ifmmode{\text{MeV}}              
                     \else{MeV{ }}\fi}                    
\def\MSbar{\relax\ifmmode\overline                        
            {\rm MS}\else{$\overline{\rm MS}${ }}\fi}     
\def\as{\relax\ifmmode \alpha_s\else{$ \alpha_s${ }}\fi}  
\def\abar{\relax\ifmmode{\bar{a}}\else{$\bar{a}${ }}\fi}  
\newlength{\tabcolf} 
\newlength{\tabcols} 
\newlength{\tabcolt} 

\begin{document}
\date{\today}
~~~~~~~~~~~~~~~~~~~~~~~~~~~~~~~~~~~~~~~~~~~~~~~~~~~~~~~~~~~~~~~
\preprint{\vbox{\hbox{\underline{\textsc{Dedicated to Prof.\ D.~V.~Shirkov's 75th birthday}}}
\hbox{\phantom{Dedicated to Prof.\ D.~V.~Shirkov's~75th~b}RUB-TPII-03/02}}}
\title{CLEO and E791 data:
       A smoking gun for the pion distribution amplitude?}

\author{Alexander~P.~Bakulev}%
 \email{bakulev@thsun1.jinr.ru}
\author{S.~V.~Mikhailov}%
 \email{mikhs@thsun1.jinr.ru}
\affiliation{%
  Bogoliubov Laboratory of Theoretical Physics,
  JINR, 141980, Moscow Region, Dubna, Russia}%

\author{N.~G.~Stefanis}
 \email{stefanis@tp2.ruhr-uni-bochum.de}
\affiliation{%
  Institut f\"ur Theoretische Physik II,
  Ruhr-Universit\"at Bochum,
  D-44780 Bochum, Germany}%
\vspace {10mm}

\begin{abstract}
The CLEO experimental data on the $\pi\gamma$ transition are analyzed
to next-to-leading order accuracy in QCD perturbation theory using
light-cone QCD sum rules.
By processing these data along the lines proposed by Schmedding and
Yakovlev, and recently revised by us, we obtain new constraints for the
Gegenbauer coefficients $a_{2}$ and $a_{4}$, as well as for the inverse
moment $\va{x^{-1}}_\pi$ of the pion distribution amplitude (DA).
The former determine the pion DA at low momentum scale, the latter is
crucial in calculating pion form factors.
From the results of our analysis we conclude that the data confirm the
end-point suppressed shape of the pion DA we previously obtained with
QCD sum rules and nonlocal condensates, while the exclusion of both
the asymptotic and the Chernyak--Zhitnitsky DAs is reinforced at the
$3\sigma$- and $4\sigma$-level, respectively.
The reliability of the main results of our updated CLEO data analysis
is demonstrated.
Our pion DA is checked against the di-jets data from the E791
experiment, providing credible evidence for our results far more
broadly.
\end{abstract}
\vspace {2mm}

\maketitle
\newpage
\section{I\lowercase{ntroduction}}
The recent high-precision CLEO results \cite{CLEO98} for the
$\pi\gamma$ transition form factor gave rise to dedicated theoretical
investigations \cite{RR96,KR96,SSK99,Kho99,SchmYa99,BMS01,DKV01,BMS02}.
These experimental data are of particular importance because they can
provide crucial quantitative information on nonperturbative parameters
of the pion DA and---as we pointed out in \cite{BMS02}---on the QCD
vacuum nonlocality parameter $\lambda_{\rm q}^{2}$, which specifies the
average virtuality of the vacuum quarks.
In the absence of a direct solution of the nonperturbative sector of
QCD, we are actually forced to extract related information from the
data, relying upon a theoretical analysis as complete and as accurate
as currently possible.

It was shown by Khodjamirian \cite{Kho99} that the light-cone QCD
sum-rule (LCSR) method provides the possibility to avoid the problem of
the photon long-distance interaction (i.e., when a photon goes on mass
shell) in the $\gamma^{*}(Q^{2})\gamma(q^{2})\to\pi^{0}$ form factor by
performing all calculations for sufficiently large $q^{2}$ and
analytically continuing the results to the limit $q^{2}=0$.
Schmedding and Yakovlev (SY) \cite{SchmYa99} applied these LCSRs to the
next-to-leading order (NLO) of QCD perturbation theory.
More recently \cite{BMS02}, we have taken up this sort of data
processing in an attempt to
(i)   account for a correct Efremov--Radyushkin--Brodsky--Lepage (ERBL)
      \cite{ERBL79} evolution of the pion DA to every measured momentum
      scale,
(ii)  estimate more precisely the contribution of the (next) twist-4
      term, and
(iii) improve the error estimates in determining the $1\sigma$- and
      $2\sigma$-error contours.

The main outcome of these theoretical analyses can be summarized as
follows:
\begin{itemize}
\item the asymptotic pion DA \cite{ERBL79} and the Chernyak--Zhitnitsky
      (CZ) \cite{CZ84} model are both outside the $2\sigma$-error
      region
\item the extracted parameters $a_{2}$ and $a_{4}$ (i.e., the
      Gegenbauer coefficients of the pion DA) are rather sensitive to
      the strong radiative corrections and to the size of the twist-4
      contribution
\item the CLEO data allow us to estimate the correlation scale
      in the QCD vacuum, $\lambda_{\rm q}^{2}$, to be
      $\lesssim 0.4$~GeV$^{2}$.
\end{itemize}

The present note gives a summary of our lengthy analysis \cite{BMS02}
extending it a few steps further, notably, by obtaining from the CLEO
data also a direct estimate for the inverse moment of the pion DA that
plays a crucial role in electromagnetic or transition form factors of
the pion and by verifying the reliability of the main results of the
CLEO data analysis quantitatively.
Moreover, we refine our error analysis by taking into account the
variation of the twist-4 contribution and treat the threshold effects
in the strong running coupling more accurately.
The predictive power of our updated analysis lies in the fact that
the value of the inverse moment obtained from an \textit{independent}
QCD sum rule is compatible with that extracted from the CLEO data,
referring in both cases to the same low-momentum scale of order of
1~GeV.
As a result, the pion DA obtained before \cite{BMS01} from QCD sum
rules with nonlocal condensates turns out to be within the
$1\sigma$-error region, while the asymptotic and the CZ pion DAs are
excluded at the $3\sigma$- and $4\sigma$-level, respectively.
Our predictions for the pion DA are found to be in agreement with the
Fermilab E791 data \cite{E79102}.

\section{L\lowercase{ight cone sum rules}}
Below, we sketch the improved NLO procedure for the data processing,
developed in \cite{BMS02}.
Let us recall that this procedure is based upon LCSRs for the
transition form factor
$F^{\gamma^*\gamma\pi}(Q^2,q^2 \approx 0)$ \cite{Kho99,SchmYa99}.
Accordingly, the main LCSR expression for the form factor
\begin{eqnarray}
 \label{eq:srggpi}
 F_\text{LCSR}^{\gamma^*\gamma\pi}(Q^2)
 = \frac1\pi\,\int\limits_0^{s_0}\!\!\frac{ds}{m_\rho^2}\,
    \rho(Q^2,s;\mu^2)
     e^{\left(m_\rho^2-s\right)/M^2}+
    \frac1\pi\,\int\limits_{s_0}^\infty\!\!\frac{ds}{s}\,
    \rho(Q^2,s;\mu^2)
\end{eqnarray}
follows from a dispersion relation.
The corresponding spectral density
$
 \rho(Q^2,s;\mu^2)\equiv\mathbf{Im}
 \left[F_\text{QCD}^{\gamma^*\gamma^*\pi}(Q^2,q^2=-s;\mu^2)\right]
$
is calculated by virtue of the factorization theorem for the form
factor at Euclidean photon virtualities
$q^2_1=-Q^2 < 0$, $q^2_2= -q^2 \leq 0$
\cite{ERBL79,DaCh81}, with $M^2\approx0.7$~GeV$^2$ being the Borel
parameter, whereas $m_\rho$ is the $\rho$-meson mass, and
$s_0=1.5$~GeV${}^2$ denotes the effective threshold in the
$\rho$-meson channel.
The factorization scale $\mu^2$ was fixed by SY at
$\mu^2=\mu^2_{\rm SY}=5.76~\gev{2}$.
Moreover, $F_\text{QCD}^{\gamma^*\gamma^*\pi}(Q^2,q^2;\mu^2)$
contains a twist-4 contribution, which is proportional to the coupling
$\delta^2(\mu^2)$, defined by \cite{Kho99,NSVVZ84}
\begin{equation}\label{eq:delta}
  \langle \pi (p)|g_{\rm s} \bar{d}
  \tilde{G}_{\alpha\mu} \gamma^{\alpha} u|0\rangle
  =i \delta^2f_{\pi} p_\mu \, ,
\end{equation}
where
$\tilde{G}_{\alpha\mu}= (1/2)\varepsilon_{\alpha\mu\rho\sigma}G^{\rho\sigma}$
and $G_{\rho\sigma}=G_{\rho\sigma}^a \lambda^a/2$.

This contribution for the asymptotic twist-4 DAs of the pion as well as
explicit expressions for the spectral density $\rho(Q^2,s;\mu^2)$
in LO have been obtained in \cite{Kho99} to which we refer for details.
The spectral density of the twist-2 part in NLO has been calculated
in \cite{SchmYa99}---see Eqs.\ (18) and (19) there.
All needed expressions for the evaluation of Eq.\ (\ref{eq:srggpi}) are
collected in the Appendix E of \cite{BMS02}, cf. Eqs.\ (E.1)--(E.3).

We set $\mu^2=Q^2$ in
$F_\text{QCD}^{\gamma^*\gamma^*\pi}(Q^2,q^2;\mu^2)$
and use the complete 2-loop expression for the form factor, absorbing
the logarithms into the coupling constant and the pion DA evolution at
the NLO level \cite{BMS02} so that
$\alpha_{\rm s}(\mu^2)
\stackrel{\rm RG}{\longrightarrow}\alpha_{\rm s}(Q^2)$
(RG denotes the renormalization group) and
\begin{eqnarray}
 \varphi_\pi(x;\mu^2)
  \stackrel{\rm ERBL}{\longrightarrow}
   \varphi_\pi(x; Q^2)=U(\mu^2 \to Q^2)\varphi_\pi(x; \mu^2). \nonumber
\end{eqnarray}
Then, we use the spectral density \hbox{$\rho(Q^2,s)$}, derived in
\cite{SchmYa99} at \hbox{$\mu^2=\mu^2_\text{SY}$}, in Eq.\
(\ref{eq:srggpi}) to obtain \hbox{$F^{\gamma^*\gamma\pi}(Q^2)$} and fit
the CLEO data over the probed momentum range, denoted by
\hbox{$\{Q^2_\text{exp}\}$}.
In our recent analysis \cite{BMS02} the evolution
$\varphi_\pi(x; Q^2)=U(\mu^2_\text{SY}
\to Q^2)\varphi_\pi(x; \mu^2_\text{SY})$
was performed \textit{for every individual point} $Q^2_\text{exp}$,
with the aim to return to the normalization scale $\mu^2_\text{SY}$
and to extract the DA parameters $(a_2,~a_4)$ at this reference
scale for the sake of comparison with the previous SY results
\cite{SchmYa99}.
Stated differently, for every measurement,
$\{Q_\text{exp}^2,F^{\gamma^*\gamma\pi}(Q_\text{exp}^2)\}$,
its own factorization and renormalization scheme was used so that the
NLO radiative corrections were taken into account in a complete way.
The accuracy of this procedure is still limited mainly owing to the
uncertainties of the twist-4 scale parameter \cite{BMS02},
$k\cdot \delta^2$, where the factor $k$ expresses the deviation of
the twist-4 DAs from their asymptotic shapes.
(Another source of uncertainty originates from the mixing of the NLO
approximations for the leading twist with the twist-4 contribution at
LO, see \cite{BMS02}.)

To summarize, the focal points of our procedure of the CLEO data
processing are
(i) $\alpha_{\rm s}(Q^2)$ is the exact solution of the 2-loop RG
    equation with the threshold $M_q=m_q$ taken at the quark mass
    $m_q$, rather than adopting the approximate popular expression in
    \cite{PDG2002} that was used in the SY analysis.
    This is particularly important in the low-energy region
    $Q^2 \sim 1$~GeV${}^{2}$, where the difference between these two
    couplings reaches about $20\%$.
(ii) All logarithms $\ln(Q^2/\mu^2)$ appearing in the coefficient
     function are absorbed into the evolution of the pion DA,
     performed separately at each experimental point $Q^2_\text{exp}$.
(iii) The value of the parameter $\delta^2$ has been re-estimated in
      \cite{BMS02} to read
      $\delta^2(1 \gev{2}) = 0.19\pm0.02~\gev{2}$.
The present study differs from the SY approach in all these points and
extends our recent analysis \cite{BMS02} with respect to points (i) and
(iii) yielding to significant improvements of the results.

It turns out that the effect of varying the shapes of the twist-4 DAs
exerts a quite strong influence which entails $k$ to deviate from 1,
i.e., from the asymptotic form.
Note that next-to-leading-order corrections in the conformal spin
\cite{BF90} for the twist-4 DAs cancel out exactly in the final
expression for $F_\text{QCD}^{\gamma^*\gamma^*\pi}$ and therefore this
deviation is due to more delicate effects.
In the absence of reliable information on higher twists, one may assume
that this uncertainty is of the same order as that for the leading-twist
case.
Therefore we set $k = 1 \pm 0.1$.
As a result, the final (rather conservative) accuracy estimate for the
twist-4 scale parameter can be expressed in terms of
\hbox{$k\cdot\delta^2(1 \gev{2}) = 0.19 \pm 0.04~\gev{2}$}, a value
close to $0.20$~GeV${}^2$ used in \cite{Kho99}.

\section{C\lowercase{onfrontation with the} CLEO \lowercase{data}}
\textbf{Pion DA vs the experimental data.}
To produce the complete $2\sigma$- and $1\sigma$-contours,
corresponding to these uncertainties, we need to unite a number of
regions, resulting from the processing of the CLEO data at different
values of the scale parameter $k\cdot\delta^2$ within this admissible
range.
This is discussed in technical detail in \cite{BMS02}.
Here, we only want to emphasize that our contours are more stretched
relative to the SY ones.
The obtained results for the asymptotic DA (\ding{117}), the BMS model
(\ding{54}) \cite{BMS01}, the CZ DA ({\footnotesize \ding{110}}),
the SY best-fit point ({\footnotesize\ding{108}}) \cite{SchmYa99},
a recent transverse lattice result ({\footnotesize\ding{116}})
\cite{Dal02}, and two instanton-based models, viz., (\ding{72})
\cite{PPRWK99} and (\ding{70}) (using in this latter case
$m_q=325$~MeV, $n=2$, and $\Lambda=1$~GeV) \cite{Pra01}, are compiled
in Table \ref{tab-1} for the maximal, middle, and minimal twist-4
scale parameter.

\begin{table}[h]
\caption{\label{tab-1}
 Models/fits for different values of $k\cdot\delta^2$
 (see text).}
\centerline{\begin{tabular}{|ccc|cc|cc|}
 \hline
 ~$k\cdot\delta^2$&
   \multicolumn{2}{c|}{$0.23$ ~GeV$^2$}
    &\multicolumn{2}{c|}{$0.19$ ~GeV$^2$}
     &\multicolumn{2}{c|}{$0.15$ ~GeV$^2$} \\ \hline
 ~Models/fits $\vphantom{^|_|}$
  & ~~~~$(a_2,a_4)\big|_{\mu_\text{SY}^2}$ & $\chi^2$
  & ~~~~$(a_2,a_4)\big|_{\mu_\text{SY}^2}$ & $\chi^2$
  & ~~~~$(a_2,a_4)\big|_{\mu_\text{SY}^2}$ & $\chi^2$
 \\ \hline
 ~best-fit  $\vphantom{^|_|}$ 
  & $(+0.28,-0.29)$               & $0.47$
  & $(+0.22,-0.22)$               & $0.47$
  & $(+0.16,-0.16)$               & $0.47$
 \\
 ~{\footnotesize\ding{108}} $\vphantom{^|_|}$ 
  & $(+0.19,-0.14)$               & $1.0$
  & $(+0.19,-0.14)$               & $0.56$
  & $(+0.19,-0.14)$               & $0.57$
 \\
~\ding{54}   $\vphantom{^|_|}$ 
  & $(+0.14,-0.09)$               & $1.7$
  & $(+0.14,-0.09)$               & $0.89$
  & $(+0.14,-0.09)$               & $0.52$
 \\
~\ding{117}  $\vphantom{^|_|}$ 
  & $(-0.003,+0.00)$              & $5.9$
  & $(-0.003,+0.00)$              & $3.9$
  & $(-0.003,+0.00)$              & $2.3$
 \\
~{\footnotesize\ding{110}} $\vphantom{^|_|}$ 
  & $(+0.40,-0.004)$              & $4.0$
  & $(+0.40,-0.004)$              & $5.2$
  & $(+0.40,-0.004)$              & $7.0$
\\
~{\footnotesize\ding{116}}  $\vphantom{^|_|}$ 
  & $(+0.06,+0.01)$              & $3.8$
  & $(+0.06,+0.01)$              & $2.3$
  & $(+0.06,+0.01)$              & $1.2$
\\
~\ding{72}   $\vphantom{^|_|}$ 
  & $(+0.03,+0.005)$              & $4.7$
  & $(+0.03,+0.005)$              & $2.9$
  & $(+0.03,+0.005)$              & $1.6$
\\
~\ding{70}   $\vphantom{^|_|}$ 
  & $(+0.06,-0.01)$               & $3.6$
  & $(+0.06,-0.01)$               & $2.1$
  & $(+0.06,-0.01)$               & $1.1$
\\ \hline
\end{tabular}
}
\end{table}
\vspace{0.5cm}

\begin{figure}[th]
 \centerline{\includegraphics[width=0.6\textwidth]{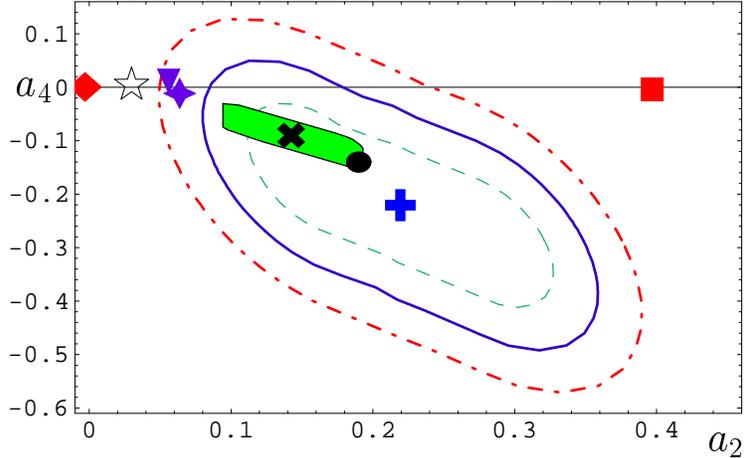}}
  \caption{\label{fig1}
   \footn
    Analysis of the CLEO data on
    $F_{\pi\gamma^{*}\gamma}(Q^2)$
    in terms of error regions around the best-fit point (\ding{58})
    (broken line: $1\sigma$; solid line:
    $2\sigma$; dashed-dotted line: $3\sigma$)
    in the ($a_2$,$a_4$) plane contrasted with various theoretical
    models explained in the text.
    The slanted shaded rectangle represents the constraints on
    ($a_2,~a_4$) posed by the nonlocal QCD sum rules
    \protect\cite{BMS01} for the value
    $\lambda^2_{\rm q}=0.4$~GeV$^{2}$.
    All constraints are evaluated at $\mu^2_\text{SY}=5.76~\gev{2}$
    after NLO ERBL evolution.}
\end{figure}

We turn now to the important topic of whether or not the set of CLEO
data is consistent with the nonlocal QCD sum-rule results for
$\varphi_{\pi}$.
We present in Fig.\ \ref{fig1} the results of the data analysis for the
twist-4 scale parameter $k\cdot \delta^2$ varied in the interval
$[0.15 \leq k \cdot \delta^2 \leq 0.23]~\gev{2}$
that includes both kinds of the discussed uncertainties of twist-4.
We have already established in \cite{BMS01} that a two-parameter model
$\varphi_{\pi}(x;a_2, a_4)$ factually enables us to fit all the moment
constraints that result from nonlocal QCD sum rules (see \cite{MR89}
for more details).
It should be stressed, however, that the restriction on the Gegenbauer
harmonics of order 2 is not just a plausible hypothesis but the direct
result of the nonlocal QCD sum-rule approach for the pion DA.
The next higher Gegenbauer harmonics up to the calculated order 10 turn
out to be too small \cite{BMS01} and are therefore neglected.
The only parameter entering the nonlocal QCD sum rules is the
correlation scale $\lambda^2_{\rm q}$ in the QCD vacuum, known from
nonperturbative calculations and lattice simulations \cite{DDM99,BM02}.
A whole ``bunch'' of admissible pion DAs resulting from the nonlocal
QCD sum-rule analysis associated with
\hbox{$\lambda_{\rm q}^2=0.4~\gev{2}$}
at \hbox{$\mu^2_0\approx1~\gev{2}$}
was determined \cite{BMS01}, with the optimal one given analytically by
$
 \varphi_{\pi}^{\text{BMS}}(x)
=
 \varphi_{\pi}^{\text{as}}(x)
      \Big[1 + a_2^{\text{opt}} \cdot C^{3/2}_2(2x-1)
     + a_4^{\text{opt}} \cdot C^{3/2}_4(2x-1)
      \Big]\,,
$
where
\hbox{$\varphi_{\pi}^{\text{as}}(x)=6x(1-x)$}
and \hbox{$a_2^{\text{opt}}=0.188$},
\hbox{$a_4^{\text{opt}}=-0.13$}
are the corresponding Gegenbauer coefficients.
From Fig.\ \ref{fig1} we observe that the nonlocal QCD sum-rule
constraints, encoded in the slanted shaded rectangle, are in rather
good overall agreement with the CLEO data at the $1\sigma$-level.
This agreement could eventually be further improved by adopting still
smaller values of $\lambda^2_{\rm q}$, say, $0.3$~GeV$^2$, which
however are not supported by the QCD sum-rule method and also lattice
calculations \cite{BM02}.
On the other hand, as it was demonstrated in \cite{BMS02}, the
agreement between QCD sum rules and CLEO data fails for larger values
of $\lambda^2_{\rm q}$, e.~g., $0.5~\gev{2}$.

 \textbf{Reliability of the main conclusions.}
The main qualitative conclusion of the presented analysis is that the
``bunch'' of pion DAs, derived in \cite{BMS01} from nonlocal QCD sum
rules, agrees rather well with the CLEO data at the $1\sigma$-level,
while both the CZ model and the asymptotic DA are ruled out at least at
the $2\sigma$-level.
However, the value of the twist-4 contribution turns out to be a subtle
point of the CLEO data processing.
Having this in mind, let us inspect the stability of the main
conclusions under the scope of the uncertainties associated with this
contribution.

We have included the twist-4 parameter $\delta^2$ in conjunction with
the vacuum quark virtuality \cite{BMS02},
$\delta^2 \approx \lambda_{\rm q}^2/2$.
Now let us ignore this relation and assume that the total twist-4
uncertainty is put by hand to an extreme uncertainty of, say, 30\%,
shifting the value of $k \cdot \delta^2$, at the low limit of the
uncertainty, to $k \cdot \delta^2 = 0.13~\gev{2}$.
Would this change our conclusions dramatically?
The result of this exercise is presented in Fig.\ \ref{fig2}(a):
\begin{figure*}[h]
 \centerline{\includegraphics[width=\textwidth]{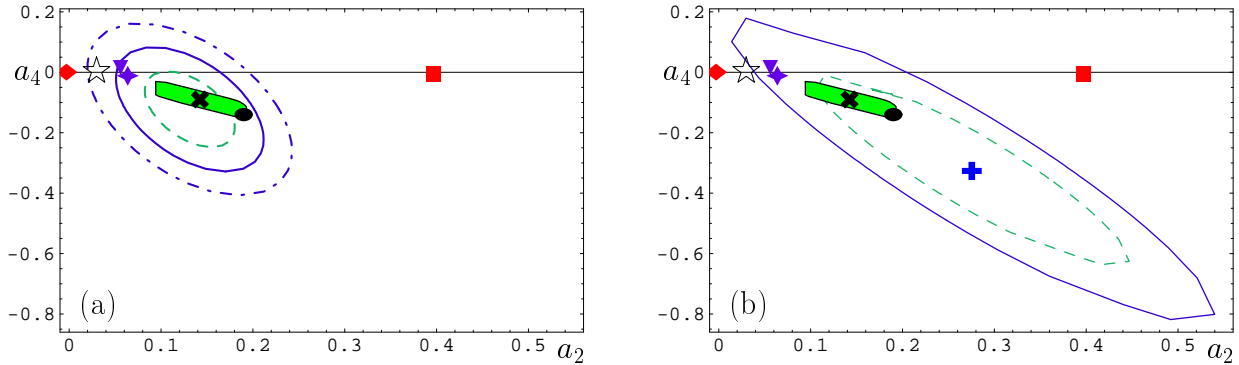}}%
  \caption{\label{fig2}
   \footn
    Analysis of the CLEO data:
    (a) Assuming a twist-4 uncertainty of 30\% or equivalently at
        $\delta^2 = 0.13~\gev{2}$.
    (b) Excluding the lowest 6 experimental points up to
        $Q^2=3~\gev{2}$.
        The designations here are the same as in Fig.\ \ref{fig1}
        and the reference scale is $\mu^2_\text{SY}=5.76~\gev{2}$.
}
\end{figure*}
One observes from Fig.\ \ref{fig2}(a) that the asymptotic DA
(\ding{117}) is still outside the $3\sigma$ error contour
(dashed-dotted line) and that the CZ point ({\footnotesize\ding{110}})
is still far-away, while instanton-based models are just at the
boundary of the $3\sigma$- (\ding{72}) or $2\sigma$- (\ding{70})
ellipse.
At the same time, the BMS\ model (\ding{54}) moved practically to the
center of the data region, whereas the Schmedding-Yakovlev best-fit
point ({\footnotesize\ding{108}}) ran outside the $1\sigma$-region.

Another way to suppress the uncertainties of the twist-4 contribution
is to repeat the processing of the CLEO data, excluding the low
momentum transfer tail.
At low $Q^2$, the twist-4 contribution strongly affects the total form
factor and, therefore, this exclusion can reduce the potential twist-4
uncertainties significantly.
To study this effect in more detail, we removed in the data processing
the lowest 6 experimental points (which possess very small errors) up
to $Q^2_{\rm exp}= 3~\gev{2}$ reducing this way the relative influence
of the twist-4 contribution by factors of magnitude.
Of course, the admissible $\sigma$-regions for the $a_{2}$, $a_{4}$
parameters become much larger now due to this exclusion, as one sees
from Fig.\ \ref{fig2}(b) in comparison with the LHS---a price one has
to pay for the restricted way of data processing.
Nevertheless, our main results and conclusions, discussed above, remain
valid with the BMS model still inside the $1\sigma$ ellipse and the
asymptotic DA outside.

\begin{figure*}[th]
 \centerline{\includegraphics[width=0.5\textwidth]{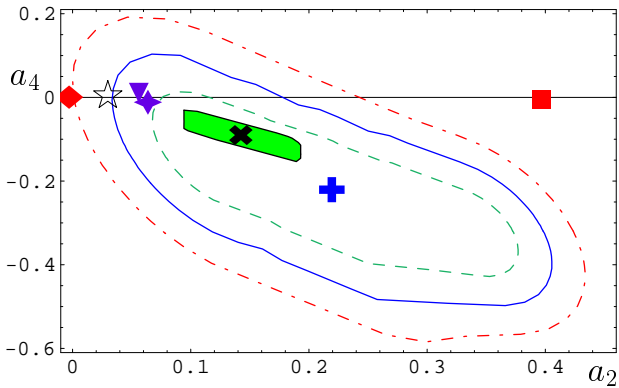}}%
  \caption{\label{fig3}
   \footn
    Estimation of the influence of higher-order corrections by varying
    the reference scale in the range $[Q^2/2, 2Q^2]$.
    The designations are as in Fig.\ \ref{fig1}.
}
\end{figure*}

The unknown high-order QCD radiative corrections provide another
important source of systematic uncertainties.
To estimate their size one should have at least the complete NNLO
coefficient function of the process.
A partial result, obtained quite recently in \cite{MMP03},
gives a hint that the size of this contribution can be large.
Therefore, the complete NNLO QCD calculation in the
$\overline{\text{MS}}$-scheme is a vital problem.
In the absence of complete results, one can only roughly estimate
the size of higher-order corrections
by varying the reference scale
$\mu^2=\mu_{\rm R}^{2}=\mu_{\rm F}^{2}=Q^{2}$,
say, in the interval $[Q^{2}/2, 2Q^{2}]$.
The corresponding ``shaking'' of the form factor is taken into account
in the systematic theoretical error that is demonstrated in
Fig.\ \ref{fig3}.
This uncertainty is rather large, of the order of $1 \sigma$, and as a
result, the set of the model predictions discussed above appears now
inside or near the $2\sigma$ contour (see Fig.\ \ref{fig3}).
Nevertheless our main conclusions remain valid.

\section{T\lowercase{he inverse moment} $\langle x^{-1}\rangle_{\pi}$
         \lowercase{vs the} CLEO \lowercase{data}}
\begin{figure*}[h]
 \centerline{\includegraphics[width=\textwidth]{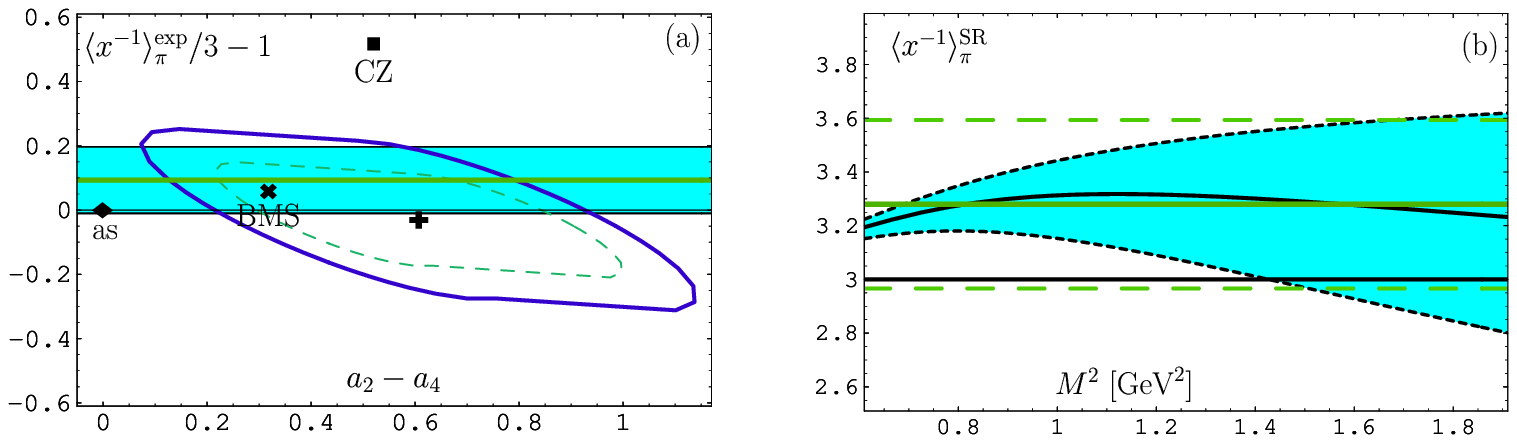}}%
  \caption{\label{fig4}
   \footn
    (a) The result of the CLEO data processing for the quantity
        $\Ds \langle x^{-1} \rangle^\text{exp}_{\pi}/3-1$ at the scale
        $\mu^2_0 \approx 1~\gev{2}$ in comparison with the
        theoretical predictions from QCD sum rules, denoted SR.
        The thick solid-line contour corresponds to the union of
        $2\sigma$-contours, while the thin dashed-line contour denotes
        the union of $1\sigma$-contours.
        The light solid line with the hatched band indicates the
        mean value of $\Ds \langle x^{-1} \rangle^\text{SR}_{\pi}/3-1$
        and its error bars in the second part of the Figure.
    (b) The inverse moment $\Ds \langle x^{-1} \rangle^\text{SR}_{\pi}$
        shown as a function of the Borel parameter $M^2$ from the
        nonlocal QCD sum rules at the same scale
        $\mu^2_0$ \protect\cite{BMS01}; the light solid line is the
        estimate for
        $\Ds \langle x^{-1} \rangle^\text{SR}_{\pi}$; finally, the
        dashed lines correspond to its error-bars.}
\end{figure*}

As already mentioned in the Introduction, in the present study we have
processed the CLEO data in such a way as to obtain an experimental
constraint on the value of the inverse moment
\hbox{$\langle x^{-1}\rangle_{\pi}(\mu^2)
=\int^1_0 \varphi_\pi(x;\mu^2){x}^{-1}dx$}
that appears in different perturbative calculations of pion form
factors.
This is illustrated in Fig.\ \ref{fig4}(a), where the positions of the
asymptotic DA, the CZ model, and the BMS one are also displayed.

Fig.\ \ref{fig4}(b) shows the theoretical estimate of the inverse
moment obtained in the framework of nonlocal QCD sum rules.
In fact, a ``daughter sum rule'' has been previously constructed
directly for this quantity by integrating the RHS of the sum rule for
$\varphi_\pi(x)$ with the weight $x^{-1}$, (for details, see
\cite{BM98,BMS01}).
Due to the smooth behavior of the nonlocal condensate at the end points
$x=0,1$, this integral is well defined eo ipso, supplying us with an
\textit{independent} QCD sum rule, with a rather good stability
behavior of $\langle x^{-1}\rangle_{\pi}^\text{SR}(M^2)$, as one sees
from this figure.
Note that we have estimated
\hbox{$\Ds \langle x^{-1}\rangle_{\pi}^\text{SR}(\mu^2_0
\approx 1~\gev{2})= 3.28 \pm 0.31$}
at the value $\lambda_{\rm q}^2=0.4~\gev{2}$ of the nonlocality
parameter.
It should be emphasized that this estimate is not related to the
model pion DA,
$\varphi^\text{BMS}_{\pi}(x;a_2,a_4)$,
constructed within the same framework.
Nevertheless, the value obtained with the ``daughter'' QCD sum rule and
those calculated using the ``bunch'' of pion DAs, mentioned above,
$\langle x^{-1}\rangle_{\pi}^\text{BMS}(\mu^2_0)
=3.17 \pm 0.09$~\cite{BMS01},
match each other.
This fact provides further support for the self-consistency of the
approach, as one appreciates by comparing the hatched strip with the
BMS point (\ding{54}) in Fig.\ \ref{fig4}(a).

It is important to notice at this point that from the CLEO data one
also obtains a constraint on the value of
\hbox{$\Ds a_2+a_4
=\langle x^{-1} \rangle^\text{exp}_{\pi}(\mu^2_0)/3-1$}
for the two Gegenbauer coefficients model.
This constraint should be compared with the independent (from the
theoretical model) estimate
$\Ds \langle x^{-1}\rangle_{\pi}^\text{SR}(\mu^2_0)$,
as mentioned above.

Let us discuss these results in more detail.
In Fig.\ \ref{fig4}(a) we demonstrate the united regions, corresponding
to the merger of the $2\sigma$-contours (solid thick line) and the
$1\sigma$-contours (thin dashed line), which have been obtained for
values of the twist-4 scale parameter within the determined range (cf.\
Table \ref{tab-1}).
This resulting admissible region is strongly stretched along the
$(a_2-a_4)$ axis, with the displayed models steered along
(approximately) the same axis, demonstrating the poor accuracy for this
combination of the DA parameters, while more restrictive constraints
are obtained for
\hbox{$\Ds \langle x^{-1} \rangle^\text{exp}_{\pi}$}.
One appreciates that the nonlocal QCD sum-rules result
$\Ds \langle x^{-1} \rangle^\text{SR}_{\pi}$,
with its error bars, appears to be in good agreement with the
constraints on
$\Ds \langle x^{-1} \rangle^\text{exp}_{\pi}$ at the $1\sigma$-level,
as one sees from the light solid line within the hatched band in
Fig.\ \ref{fig4}(a).
In particular, the $1\sigma$-constraint obtained at the central value
$k\cdot \delta^2=0.19~\gev{2}$
exhibits the same good agreement with the corresponding sum-rule
estimate because the theoretical uncertainty of the twist-4 scale
parameter and of the radiative correction, already mentioned, affect
mainly the $(a_{2}-a_{4})$ constraint.
The CLEO best-fit point (\ding{58}) in Fig. \ref{fig4}(a) is near to
zero in accordance with the previous data-processing results,
presented in the first line of Table 1, $a_2+a_4\simeq 0$.
Moreover, the estimate $\Ds \langle x^{-1} \rangle^\text{SR}_{\pi}$
is close to
$\Ds \langle x^{-1} \rangle^\text{EM}_{\pi}/3-1=0.24\pm 0.16$,
obtained in the data analysis of the electromagnetic pion form factor
within the framework of a different LCSR method in \cite{BKM00,BK02}.
These three independent estimates are in good agreement to each other,
giving robust support that the CLEO data processing, on one hand, and
the theoretical calculations, on the other, are mutually consistent.
Moreover, Dorokhov~\footnote{Private communication.} recently
obtained from the instanton-induced effective theory model
$\varphi_{\pi}^\text{I-mod}(x)$~\cite{Dor02} the estimate
$\Ds \langle x^{-1} \rangle_{\pi}^\text{I-mod}/3-1\approx -0.09$,
which is close to the CLEO result.

More importantly, the end-point contributions to the
$\langle x^{-1}\rangle_{\pi}^\text{SR}$
are suppressed, the range of suppression being controlled by the value
of the parameter $\lambda_{\rm q}^2$.
The larger this parameter, at fixed resolution scale
$M^{2} > \lambda_{\rm q}^2$,
the stronger the suppression of the nonlocal-condensate contribution.
Similarly, an excess of the value of
$\langle x^{-1}\rangle_{\pi}$ over 3 (asymptotic DA) is also
controlled by the value of $\lambda_{\rm q}^2$, becoming smaller with
increasing $\lambda_{\rm q}^2$.
Therefore, to match the value
$\langle x^{-1}\rangle_{\pi}^\text{SR}$
to the CLEO best-fit point, would ask to use larger values of
$\lambda_{\rm q}^2$ than $0.4~\gev{2}$.
But this is in breach of the $(a_2,a_4)$ error ellipses.
A window of about $0.05$~GeV${}^2$ exists to vary $\lambda_{\rm q}^2$:
any smaller and one is at the odds with QCD sum rules and lattice
calculations \cite{BM02}; any larger and the nonlocal QCD sum-rules
rectangle can tumble out of the CLEO data region.

\section{C\lowercase{omparison with the} E791 \lowercase{data}}
Very recently, an independent source of experimental data by the E791
Fermilab experiment \cite{E79102} has become available providing
additional constraints on the shape of the pion DA.
However, these data are affected by inherent uncertainties and their
theoretical explanation by different groups \cite{Che01,NSS01,BISS02}
is controversial.
It is not our goal here to improve the theoretical framework for the
calculation of diffractive di-jets diffraction.
For our purposes it suffices to show basically two things: first, that
our predictions for this process are not conflicting the E791 data and
second, to show \textit{in comparison} with other models for the pion
DA that the BMS model has best agreement with these data, using for all
considered models the \textit{same} calculational framework.

\begin{figure}[ht]
 \centerline{\includegraphics[width=0.6\textwidth]{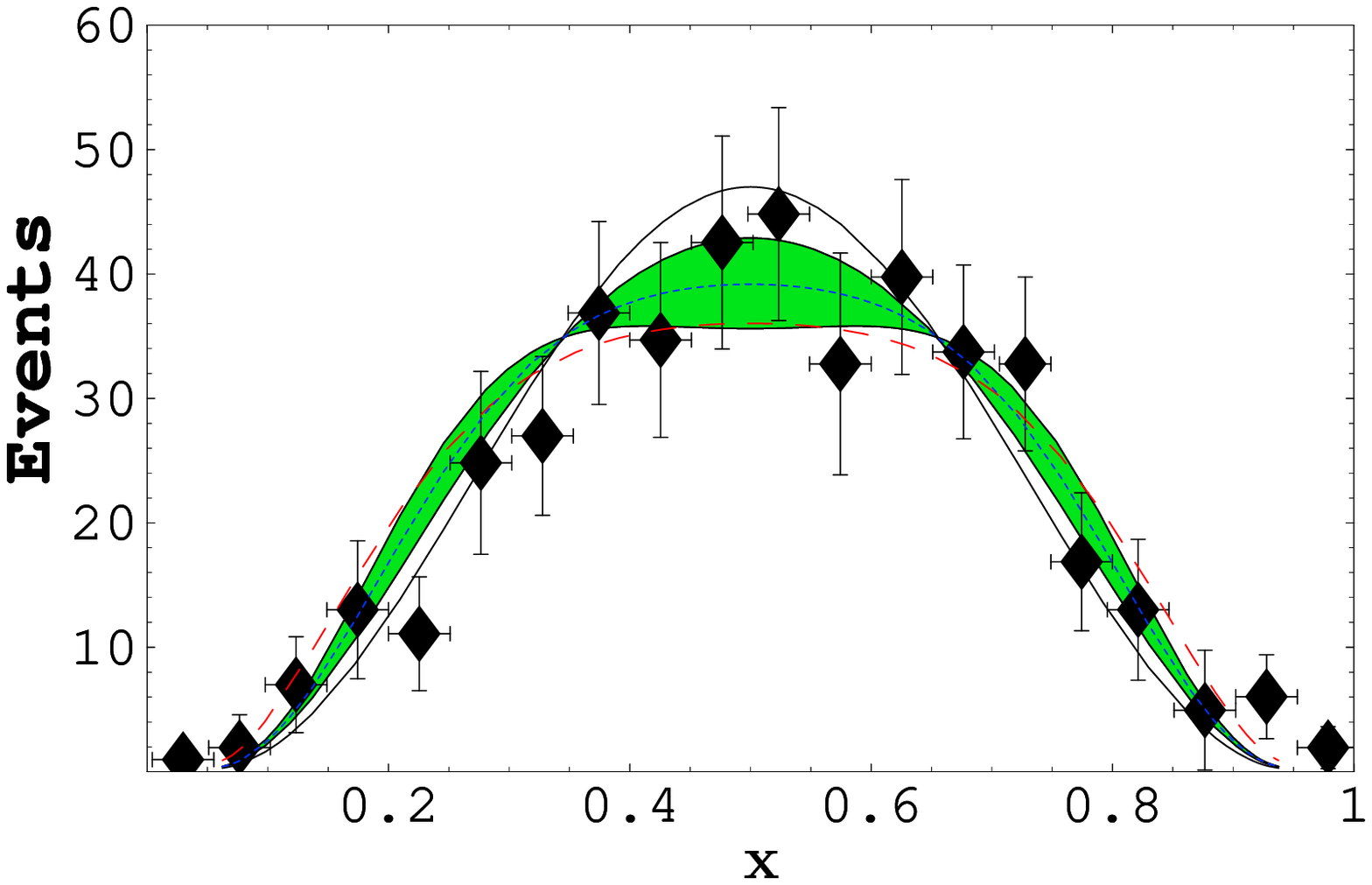}}
  \caption{\label{fig5}
   \footn
   Comparison of $\varphi^\text{as}$ (solid line),
   $\varphi^\text{CZ}$ (dashed line), and the
   BMS ``bunch'' of pion DAs (strip, \protect\cite{BMS02})
   with the E791 data \protect\cite{E79102}.
   The corresponding $\chi^2$ values are As: 12.56; CZ: 14.15;
   BMS: 10.96.}
\end{figure}

To compare our model DA for the pion \cite{BMS01} with the E791 di-jet
events \cite{E79102} and other pion DAs, we adopt the convolution
approach developed in \cite{BISS02} having also recourse to
\cite{{FrMcD02}}.
The results of the calculation are displayed in Fig.\ \ref{fig5} making
evident that, though the data from E791 are not that sensitive as to
exclude other shapes for the pion DA (asymptotic and CZ model), also
displayed for the sake of comparison, they are relatively in good
agreement with our prediction.
Especially, in the middle $x$ region, where our DAs ``bunch'' has the
largest uncertainties (see \cite{BMS01}), the predictions are not in
conflict with the data.
Note, however, that all theoretical predictions shown in Fig.\
\ref{fig5} are not corrected for the detector acceptance.
For a more precise comparison, this distortion must be taken into
account.

\section{C\lowercase{onclusions}}
Let us summarize our findings.
They have been obtained by refining the CLEO data analysis, we
initiated in \cite{BMS02}, in the following points.
We corrected for the mass thresholds in the running strong coupling
and incorporated the variation of the twist-4 contribution more
properly.
In addition, the CLEO data were used to extract a direct constraint
on the inverse moment
$\langle x^{-1}\rangle_{\pi}(\mu^2_0)$
of the pion DA---at the core of form-factor calculations.
This has relegated the CZ model and the asymptotic pion DAs beyond, at
least, the $3\sigma$-level (confidence level of 99.7\%), with the SY
best-fit point still belonging to the $1\sigma$ deviation region
(68\%) in the parameter space of $(a_{2},a_{4})$, while providing
compelling argument in favor of our model \cite{BMS01}, which is
also within this error ellipse and remains there even assuming
a potentially higher twist-uncertainty of the order of 30\%.

Both analyzed experimental data sets (CLEO \cite{CLEO98} and Fermilab
E791 \cite{E79102}) converge to the conclusion that the pion DA is not
everywhere a convex function, like the asymptotic one, but has instead
two maxima with the end points ($x=0,1$) strongly suppressed---in
contrast to the CZ DA.
These two key dynamical features of the DA are both controlled by the
QCD vacuum inverse correlation length $\lambda_{\rm q}^2$, whose value
suggested by the CLEO data analysis here and in \cite{BMS02} is
approximately 0.4~GeV$^2$ in good compliance with the QCD sum-rule
estimates and lattice computations \cite{BM02}.
\bigskip

\noindent \textbf{Acknowledgments.}
We are grateful to D.~Ashery, D.~Ivanov, A.~Khodjamirian, N.~Nikolaev,
M.~Polyakov, and O.~Teryaev for discussions and useful remarks.
Two of us (A.~B. and M.~S.) are indebted to Prof.\ Klaus Goeke
for the warm hospitality at Bochum University, where this work was
partially carried out.
This work was supported in part by INTAS-CALL 2000 N 587, the RFBR
(grant 03-03-16816), the Heisenberg--Landau Programme (grants 2002
and 2003), the COSY Forschungsprojekt J\"ulich/Bochum, and the Deutsche
Forschungsgemeinschaft (DFG).

\end{document}